\documentclass[%
 reprint,
 amsmath,amssymb,
 aps,
]{revtex4-2}

\usepackage{graphicx}% Include figure files
\usepackage{dcolumn}% Align table columns on decimal point
\usepackage{bm}% bold math
\begin{document}

\preprint{APS/123-QED}

\title{Medium modified Fragmentation Functions \\ with open source xFitter}% Force line breaks with \\
%\thanks{A footnote to the article title}%

\author{P\'ia Zurita}
% \homepage{http://www.Second.institution.edu/~Charlie.Author}
\affiliation{Institut f\"ur Theoretische Physik, Universit\"at Regensburg, 93040 Regensburg, Germany}%
%\affiliation{
% Third institution, the second for Charlie Author
%}%

\date{\today}% It is always \today, today,
             %  but any date may be explicitly specified

\begin{abstract}
A new analysis of the modification of the hadronization process in the nuclear medium for pions is presented. The effective description is condensed in a %novel 
set of medium modified fragmentation functions (nFFs) obtained at next-to-leading order (NLO) accuracy. The study was made using the open-source tool xFitter, conveniently modified to incorporate semi-inclusive processes. Theoretical uncertainties for the nFFs are also provided.
%\begin{description}
%\item[Usage]
%Secondary publications and information retrieval purposes.
%\item[Structure]
%You may use the \texttt{description} environment to structure your abstract;
%use the optional argument of the \verb+\item+ command to give the category of each item. 
%\end{description}
\end{abstract}

%\keywords{Suggested keywords}%Use showkeys class option if keyword
                              %display desired
\maketitle

\tableofcontents

\section{\label{sec:intro} Introduction}

In the perturbative QCD (pQCD) description of scattering processes the initial and final state hadrons involved are understood in terms of parton distribution functions (PDFs) and fragmentation functions (FFs), respectively. The former are quite well constrained from existing data from deeply inelastic scattering (DIS) and several observables in p+p collisions up to next-to-next-to leading order (NNLO) accuracy, while the latter, not as well determined, are know up to NNLO in the case of semi-inclusive $l^{+}+l^{-}$ annihilation (SIA) and up to NLO from SIA, semi-inclusive DIS (SIDIS) and single particle production in p+p collisions. 

When an experiment is performed using nuclei a plethora of different phenomena arise, phenomena that can not be explained by naively considering the nucleus as a collection of non-interacting nucleons. In the case of inclusive DIS in the collinear factorized approach, the idea of universal medium modified or nuclear PDF (nPDF) was introduced with success to describe the observed measurements. Obtained through global fits to the world data \cite{Eskola:2016oht,Kusina:2020lyz,AbdulKhalek:2020yuc,Walt:2019slu,Khanpour:2020zyu,deFlorian:2011fp} or through theoretical modelling\cite{Kulagin:2014vsa}, nPDFs can describe within uncertainties many of the features observed in DIS off nuclear targets and in p(d)+A collisions. They are not as well known as free proton PDFs due to several reasons (e.g. the precision and reduced kinematical coverage of the data), but the situation is expected to improve with results from ongoing and future experiments \cite{Paukkunen:2020rnb,Aschenauer:2017oxs}.

%among several reasons (e.g. reduced kinematical coverage of the data), but the situation is expected to improve with results from ongoing and future experiments.

In the case of the final state, however, much remains to be discovered. For decades it has been known that the production of hadrons in a nuclear medium deviates non trivially from the case with only protons. Interactions between the partons with the medium before hadronization, interactions between the formed hadrons and the medium, and in-medium modification of the evolution equations are some of the mechanisms proposed to explain the observed data \cite{Arleo:2008dn,Accardi:2009qv,Li:2020zbk}. Despite being very different approaches, most of these successfully reproduce some features of the measured quantities.

While of great interest due to the rich physics under study, proton-nucleus (p+A) and nucleus-nucleus (A'+A) collisions are not the cleanest form of accessing FFs. This is specially true in the case of heavy-ion (HI) interactions where signatures of a new state of matter, the Quark Gluon Plasma (QGP), could be mixed with the so called cold nuclear matter effects. Thus, the simplest way of studying the modification of the FFs in the nuclear medium is to observe SIDIS with a nuclear beam/target. For charged hadrons non trivial deviations from the vacuum case have been found \cite{Ashman:1991cx,Osborne:1978ai,Adams:1993mu}. However, to date only the fixed target experiment HERMES has provided production rates of identified hadrons in SIDIS off different nuclei \cite{Airapetian:2007vu}. Preliminary data from CLAS for $\pi^{+}$ do exist \cite{Hakobyan:2008kua}; at the time of writing the final analysis is being performed. 

In view of the success of nPDFs, it is reasonable to wonder whether the factorization can be extended to final state nuclear effects or if, and where, it breaks down. Under the same assumptions used for nPDFs (validity of factorization and universality), a pioneering study was done in \cite{Sassot:2009sh}. Using a set of vacuum FFs, nuclear FFs (nFFs) were determined in a global fit to the HERMES SIDIS and RHIC single hadroproduction data. Since then (n)PDFs and vacuum FFs have been considerably improved with more precise measurements form different experiments. With future colliders that will further explore the physics of in-medium hadronization around the corner, it is timely to update the results. In the present work a new extraction of nFFs, from here on called \emph{LIKEn21}, is presented. The determination is done using an extension of the xFitter PDF tool \cite{Alekhin:2014irh,Zenaiev:2016jnq,Bertone:2017tig,xfitterlink}, built on the modification of \cite{Walt:2019slu} for nPDFs. This extension could also be used in the future to perform joint PDFs+FFs analyses, which have been shown to be an alternative form to constrain some partonic densities \cite{Sato:2019yez,Aschenauer:2019kzf}.

This article is organized as follows:
the next section describes the theoretical framework and the vacuum FFs used as baseline, together with the modifications introduced for the nFFs. Sec. \ref{sec:analysis} contains information about the changes in xFitter, the fitting procedure and the determination of uncertainties. The experimental data used in this work is discussed in Sec. \ref{sec:data} and the results are the focus of Sec. \ref{sec:results}. A summary of the analysis is given in Sec. \ref{sec:summary}. 

\section{\label{sec:framework}Theoretical framework}

\subsection{\label{sec:sidis}SIDIS and hadroproduction}

In the first part of this analysis unpolarised SIDIS off nuclear targets is considered. For an incoming/target hadron of momentum $p_{N}$ and outgoing hadron of momentum $p_{h}$, the unpolarized cross-section can be written as 
\begin{eqnarray}
\frac{d^{3}\sigma^h}{dxdydz}&=\frac{2\pi\alpha_{em}^{2}}{Q^{2}}\frac{1+(1-y)^2}{y}\Big[2F_{1}^h(x,z,Q^{2}) \nonumber\\
&+\frac{2(1-y)}{1+(1-y)^2}F_{L}^{h}(x,z,Q^{2})\Big] \, ,
\label{eq:sidis}
\end{eqnarray}
where $x$ and $y$ are the DIS scaling variables ($Q^{2}=sxy$), $z\equiv p_{h}\cdot p_{N}/p_{N}\cdot q$ and $-q^{2}=Q^{2}$. The structure functions are given, up to NLO, by
\begin{eqnarray}
2F_{1}^{h}(x,z,Q^{2})&=\sum_{q,\bar{q}} e_{q}^{2}\Big\{f_{q}(x,Q^{2})D_{q}^{h}(z,Q^{2})\nonumber \\
&+\frac{\alpha_{s}(Q^{2}}{2\pi}\Big[f_{q}\otimes C_{qq}^{1}\otimes D_{q}^{h}\nonumber \\
&+f_{q} \otimes C_{gq}^{1}\otimes D_{g}^{h}\nonumber\\
&+f_{g}\otimes C_{qg}^{1}\otimes D_{q}^{q}\Big](x,z,Q^{2}) \Big\} \, ,
\label{eq:sf1}
\end{eqnarray}
and
\begin{eqnarray}
F_{L}^{h}(x,z,Q^{2})&=\frac{\alpha_{s}(Q^{2}}{2\pi}\sum_{q,\bar{q}}e_{q}^{2}\Big[f_{q}\otimes C_{qq}^{L}\otimes D_{q}^{h}\nonumber \\
&+f_{q} \otimes C_{gq}^{L}\otimes D_{g}^{h}\nonumber\\
&+f_{g}\otimes C_{qg}^{L}\otimes D_{q}^{q}\Big](x,z,Q^{2}) \, ,
\label{eq:sfl}
\end{eqnarray}
with $f_{i}$ and $D_{i}^{h}$ the PDF and FF for parton $i$, respectively. The NLO expressions for the coefficients $C_{ij}^{1,L}$ in the $\overline{\text{MS}}$ scheme used in the present work can be found in \cite{deFlorian:1997zj}. 

In a second part of the analysis, high transverse momentum ($p_T$) charged and neutral pion production in d+Au was studied. Up to $1/p_{T}^{n}$ corrections, the differential cross-section can be written as
\begin{eqnarray}
E\frac{d^{3}\sigma^{h}}{dp^{3}}=\sum_{i,j,k} f_{i}\otimes f_{j}\otimes d\sigma^{k}_{ij}\otimes D_{k}^{h} \, ,
\label{eq:hadroprod}
\end{eqnarray}
where the sum is performed over all contributing channels $a+b\to c+X$. The partonic cross-section is known up to NLO, see e.g. \cite{Aversa:1988vb,Jager:2002xm}. And the calculations were done using the same code as in \cite{Sassot:2010bh}

\subsection{\label{sec:nffs}Nuclear FFs}

The LIKEn21 set is based on the vacuum FFs from DEHSS \cite{deFlorian:2014xna}. The baseline parametrization for the FFs is given by
\begin{eqnarray}
D_{i}^{h}(z,Q_{0})=\tilde{N}_{i}z^{\alpha_{i}}(1-z)^{\beta_{i}}\Big[1+\gamma_{i}(1-z)^{\delta_{i}}\Big] \, ,
\label{eq:dehss}
\end{eqnarray}
where $\tilde{N}$ simply indicates a different notation for the normalization than in the original paper. The initial scale is chosen to be $Q_{0}=1\text{ GeV}$ for the light quarks and the gluon, and $Q_{0}=m_{c}=1.43\text{ GeV}$ and $Q_{0}=m_{b}=4.3\text{ GeV}$ for charm and bottom, respectively. Below the corresponding thresholds $D_{c,b}^{h}=0$. Several treatments (schemes) for the heavy quarks and their masses are routinely implemented in the description of DIS and consequently extraction of PDFs but they are not widely used in the determination of FFs. While shown to be relevant \cite{Epele:2018ewr}, mass effects are completely neglected in this work in order to be consistent with the baseline FFs.

In \cite{Sassot:2009sh} the nuclear modification was introduced as a flexible weight factor $W_{i}^{h}$ in the form of an Euler Beta-function convoluted with the NLO pion FFs from \cite{deFlorian:2007aj}, a natural extension in Mellin space \cite{deFlorian:2003qf}. As the present work is not done in Mellin space such a convolution becomes numerically cumbersome. Therefore the proposed ansatz is to modify the vacuum parameters, providing them with a dependence on the mass number $A$. Namely   
\begin{eqnarray}
\tilde{N}_{i}&\to&\tilde{N}_{i}\Big[1+N_{i,1}(1-A^{N_{i,2}})\Big] \nonumber \\
c_{i}&\to&c_{i}+c_{i,1}(1-A^{c_{i,2}}) \, ,
\label{eq:params}
\end{eqnarray}
with $i$ the different partons and $c=\alpha, \beta, \gamma, \delta$. Notice that the slightly different form for the normalization is irrelevant. This type of parametrization for the nuclear effects is also used in some nPDFs analyses \cite{Walt:2019slu,Kusina:2020lyz} and has the advantage of recovering the vacuum FFs when $A=1$. 
There are only a few differences in the present work with respect to the one in DEHSS. The PDF set chosen is MMHT2014 \cite{Harland-Lang:2014zoa} instead of MSTW2008 \cite{Martin:2009iq}, which should not have a significant impact. Likewise, the mass of the bottom quark was taken to be $m_{b}=4.5\text{ GeV}$, an intermediate value between the ones in MMHT2014 ($m_{b}=4.75\text{ GeV}$) and DEHSS ($m_{b}=4.3\text{ GeV}$). 
At variance with \cite{Sassot:2009sh}, where different sets of nPDFs were used for comparison, here no nuclear modification on the initial distribution was considered. The reason is twofold. On the one hand most proton PDFs already contain some nuclear information from charged-current DIS and fixed target experiments off light nuclei, so using nPDFs would effectively be double counting. On the other hand the SIDIS data considered in the fit are given as a double ratio that cancels out almost all of the initial state effects, as can be seen in Figs. 2 and 10 of \cite{Sassot:2009sh}.

\section{\label{sec:analysis}Modus operandi}

\subsection{\label{sec:xfitter}SIA and SIDIS in xFitter}

The xFitter project provides an open-source tool \cite{Alekhin:2014irh,Bertone:2017tig,Zenaiev:2016jnq,xfitterlink} to fit proton PDFs using different theoretical assumptions, up to NNLO accuracy. Most parametrization forms, several mass schemes and common evolution, computational and PDFs programs (MINUIT \cite{James:1975dr,Lazzaro:2010zza}, QCDNUM \cite{Botje:2010ay}, APFEL \cite{Bertone:2013vaa}, APPLGrid \cite{Carli:2010rw}, LHAPDF \cite{Buckley:2014ana}, etc) are included. Given its availability to any user, significant extensions have been incorporated, such as dipole models \cite{Luszczak:2013rxa,Luszczak:2016bxd} and small-x resummation \cite{Ball:2017otu,Abdolmaleki:2018jln}. In particular it has been recently expanded to include nuclear effects in the PDFs \cite{Walt:2019slu}, which could eventually be used for a joint analysis of proton and nuclear PDFs. Full details on the modifications introduced can be found in the corresponding release.

The modifications incorporated for this work build and expand upon those done in \cite{Walt:2019slu}. The main changes are the inclusion of SIA and SIDIS NLO routines and (n)FFs parametrizations, together with kinematic cuts. The running mode has been expanded to fit: (n)PDFs only (with or without SIDIS data), (n)FFs only, or both together. When fitting only (n)FFs, the (n)PDFs can be taken from the xFitter sets or directly from LHAPDF. If only PDFs are fitted, none of the FFs related routines are called. To accommodate for the vacuum and nuclear FFs for pions and kaons, the number of internal parameters in MINUIT was increased.

An extra call in the evolution is now needed for the FFs. The default routine in xFitter is QCDNUM which allows for the simultaneous evolution of PDFs and FFs. The latest version of QCDNUM \cite{Botje:2010ay} was used in this work which permits to consider an intrinsic treatment of the heavy-quarks, appropriate for DEHSS. In that mode, the charm and bottom FFs are taken to be fixed below their respective masses and evolve only for values of $Q\geq m_{c,b}$. This is crucial to implement the charm and bottom FFs in the style of DEHSS, and one only needs to set the heavy quark contributions to zero in the cross-section calculations when $Q<m_{c,b}$. The effect of evolving the PDFs with QCDNUM instead of LHAPDF is a change of $0.16\%$ in the final $\chi^{2}$. A larger discrepancy ($+2.89\%$) was found when taking $\alpha_{s}$ from LHAPDF. 
%Furthermore, it has no impact on the PDFs, as the sum rules have to be externally incorporated. 

The modifications have been tested by comparing with the pion FFs from DEHSS and DSS, finding a difference well below $1\%$ (and mostly below $0.1\%$) for all the $z$ and $Q^{2}$ values explored. A comparison with the FFs from AKK \cite{Albino:2008fy} was not done as they require to start the heavy-quark contribution at a scale different than the masses, which is not possible with QCDNUM. Instead the APFEL \cite{Bertone:2013vaa} evolution package should be used. This is left for future work.

At variance with DIS, where only one convolution over $x$ is needed, SIDIS processes require one extra integration, which slows down the computation and would render it impractical for fitting purposes. To avoid this issue the following strategy was used: in the case of fitting one type of distribution a grid containing the convolution of the hard coefficients with the non-fitted distributions is created at the first call. For the current data set the generation of the grid requires a few seconds and makes the code runs 20 times faster than without it.

\subsection{\label{sec:fitting}Fitting procedure and uncertainties}

The values of the parameters that best describe the data are obtained by minimizing a $\chi^{2}$ quantity, which can be written as
\begin{eqnarray}
\chi^{2}=\sum_{i}\frac{(m^{i}-\sum_{\alpha}\Gamma_{\alpha}^{i}b_{\alpha}-\mu^{i})^{2}}{\Delta_{i}^{2}}+\sum_{\alpha}b_{\alpha}^{2} \, ,
\label{eq:chi2_complicated}
\end{eqnarray}
with $\mu^{i}$ the measured value for point $i$, $m^{i}$ the computed value that depends on the fit parameters, and $\Delta^{i}$ the total uncorrelated uncertainties (statistical and systematic) added in quadrature. The correlated uncertainty for data point $i$ from source $\alpha$ is given by $\Gamma^{i}_{\alpha}$, while $b_{\alpha}$ are the nuisance parameters quantifying the strength of the error source. In the present case only uncorrelated uncertainties are provided, simplifying the function to minimize to 
\begin{eqnarray}
\chi^{2}=\sum_{i}\frac{(m^{i}-\mu^{i})^{2}}{\Delta_{i}^{2}} \, .
\label{eq:chi2_real}
\end{eqnarray}

In xFitter the relative uncertainties ($\delta^{i}$) are used and several options are available for the scaling of the uncertainties. The statistical ones can be treated as additive ($\Delta^{i}=\delta^{i}\mu^{i}$) or Poisson ($\Delta^{i}=\delta^{i}\sqrt{m^{i}\mu^{i}}$), the latter more appropriate when the statistical uncertainties scale with the square root of the expected number of events. With regard to the systematic uncertainties, these can be taken as multiplicative ($\Delta^{i}=\delta^{i}m^{i}$) or additive. If $\Delta^{syst}$ are proportional to the central values the former option is more adequate. In the current work a Poisson scaling for the statistical uncertainties and a multiplicative one for the systematic errors were used base on the fact that the former is used for HERA data analysis and that the systematic uncertainties are quoted to be a scale uncertainty $~3\%$. All results shown here were obtained under those assumptions. A fit with no scaling was also performed and the corresponding results are discussed in Subsec.\ref{sec:results_nffs}. 

Once the minimization of $\chi^{2}$ is achieved, a detailed study of the precision of the fit must be done. That is, one has to provide a quantitative estimation of how well the extracted parameters were determined or, in other words, how much those values can vary without spoiling the quality of the fit. Computing the observables with different sets of parameters \emph{close} to the best fit ones results in a theoretical uncertainty band, routinely given with the (n)PDFs sets and, recently, also with the FFs. 

Three methods can be used to determine the theoretical errors. The first one is the Lagrange Multiplier (LM) technique \cite{Pumplin:2000vx,Stump:2001gu,Martin:2002aw}, very robust when parameters are loosely constrained by the data. However it is computationally very demanding and the use of the error in the distributions to estimate the uncertainties of physical observables is not trivial \cite{Epele:2012vg,Martin:2002aw}, which makes it the favoured approach only for very specific studies. 

A second method is the Monte Carlo (MC) approach. In it the data are randomly varied according to the experimental uncertainties and for each variation (called MC replica) a fit is performed. At the end of the procedure one has as many sets of parton distributions as replicas ($\sim 10^2-10^3$) from which the central value and uncertainties for both (n)PDFs/FFs and observables can be derived. This method is also computationally demanding but has the advantage of being far less biased with respect to the original parametrization of the distributions. 

A third option, most favoured for PDFs, is the Hessian method \cite{Pumplin:2001ct,Pumplin:2000vx}. In it, it is assumed that the quadratic expansion is a good approximation to the $\chi^{2}$ function around the minimum 
\begin{eqnarray}
\chi^{2}\approx \chi_{0}^{2}+\sum_{i,j} (a_{i}-a_{i}^{0})H_{ij}(a_{j}-a_{j}^{0}) \, ,
\label{eq:chi2_hessian}
\end{eqnarray}
where $\chi^{2}_{0}$ is the value for the best fit, $a_{i}$ are the parameters, and $a_{i}^{0}$ their best fit values. The Hessian matrix, to be determined numerically, is:
\begin{eqnarray}
H_{ij}= \frac{1}{2}\frac{\partial^{2} \chi^{2} }{\partial a_{i}\partial a_{j}}\vert_{a_{i}=a_{i}^{0},a_{j}=a_{j}^{0}}  \, .
\label{eq:hessian}
\end{eqnarray}
As $H_{ij}$ is symmetric it can be written in terms of a complete set of orthonormal eigenvectors ($v_{ij}$) and corresponding eigenvalues ($\epsilon_{j}$). Using these, one can write the displacement of the parameters around the best fit value as
\begin{eqnarray}
a_{i}-a_{i}^{0}=\sum_{j}\frac{v_{ij}}{\sqrt{\epsilon_{j}}}z_{j}  \, ,
\label{eq:displacement}
\end{eqnarray}
with $z_{i}$ the new parameters. Clearly the parameters that receive contributions from eigenvectors with large eigenvalues will be better determined. Then, the distance between the $\chi^{2}$ and its best value is:
\begin{eqnarray}
\Delta\chi^{2}\equiv\chi^{2}-\chi_{0}^{2}\approx\sum_{i} z_{i}^{2} \, ,
\label{eq:chi2_hessian2}
\end{eqnarray}
where $\Delta\chi^{2}$ is called \emph{tolerance}. The eigensets $S^{\pm}_{i}$ are then defined as the distributions obtained by moving up and down each $z_{i}$ individually so that the total increase in the $\chi^{2}$ is equal to the tolerance. Using the error sets and the fact that the new parameters are independent among each other, the error of any quantity depending $\mathcal{O}$ on the partonic densities takes the form
\begin{eqnarray}
\mathcal{O}^{\pm}=\sqrt{\sum_{i}\text{}^{max}_{min}\Big[\mathcal{O}^{+}_{i}-\mathcal{O}_{0},\mathcal{O}^{-}_{i}-\mathcal{O}_{0},0 \Big]^{2} } \, ,
\label{eq:error_prop}
\end{eqnarray}
with $\mathcal{O}^{\pm}_{i}$ the quantity computed using eigenset $S_{i}^{\pm}$ and $\mathcal{O}_{0}$ using the central set $S_{0}$.

The adequacy of the Hessian method application to FFs was proven in \cite{Epele:2012vg} by validating it with the results from the LM technique. It is also the method used in DEHSS and therefore the one adopted in this work. The Hessian and the MC methods are both fully implemented in xFitter. 

The only remaining task is to decide how large the tolerance should be. In principle, if one were dealing with one single experiment and one single target, one should choose $\Delta\chi^{2}=1$. However fits usually deal with data from very diverse experimental and analysis conditions. In such a situation, while there is consensus on choosing a tolerance well above $1$, the criteria used to assign a numerical value greatly vary from fit to fit. In this work the same method as in the baseline vacuum FFs is employed: using the Gaussian probability density function for a $\chi^2$ distribution with $n$ degrees of freedom (d.o.f),
\begin{eqnarray}
P_{n}(\chi^{2})=\frac{(\chi^{2})^{n/2-1}e^{-\chi^{2}/2}}{\Gamma(n/2)2^{n/2}} \, ,
\label{eq:tolerance}
\end{eqnarray}
the $\Delta\chi^{2}$ corresponding to the 68th and 90th percentiles ($68\%$ and $90\%$ confidence levels, (C.L.)) were found. For the current fit the $68\%$ C.L. ($90\%$ C.L.) corresponds to $\Delta\chi^{2}=11$ ($\Delta\chi^{2}=32$). 
%%%%%%%%%%%%%%%%%%%%%%%%%%%%%%%%%%%%%%%%%%%%%%%%%

\section{\label{sec:data}Experimental data}

Unlike FFs that can be extracted from SIA data complemented with SIDIS for flavour separation, nFFs can only be derived from SIDIS and p(d)+A collisions. This work is divided in two steps: in the first one, charged and neutral pion production data from SIDIS off nuclei were used to extract the medium modification of the FFs. In a second step preliminary $\pi^{+}$ SIDIS data from CLAS \cite{Hakobyan:2008kua} and single pion production data from RHIC \cite{Abelev:2009hx,Adams:2006nd,Adare:2013esx} were compared with theoretical predictions using the nFFs previously obtained.

\subsection{\label{sec:sidisdata}SIDIS}

For more than four decades measurements of charged hadrons in SIDIS experiments off nuclei have been performed \cite{Ashman:1991cx,Osborne:1978ai,Adams:1993mu}. A detailed separation into different hadronic species is however relative novel \cite{Airapetian:2007vu}. In that work, the HERMES collaboration studied the modification of SIDIS using deuterium d, helium $^{4}$He, neon Ne, krypton Kr, and xenon Xe targets, with pions ($\pi^{+}$, $\pi^{-}$, $\pi^{0}$), kaons ($K^{+}$, $K^{-}$), protons and anti-protons identified. Moreover the data are published as distributions in the kinematical variables $z$, $Q^{2}$ and the virtual photon energy $\nu=Q^{2}/(2Mx)$ with $M$ being the mass of the nucleon. The dependence on the transverse momentum of the outgoing hadron, relevant for TMDs studies is also given. In order to exclude nucleon resonances, the constraint $W^{2}=\sqrt{2M\nu+M^{2}-Q^{2}} \geq 4\text{ GeV}^{2}$ was imposed, roughly three times lower than the usual cut in PDFs analyses ($\sim 12 \text{ GeV}^{2}$). Increasing the cut to match the one from the proton PDF would remove from the fit the $24$ data points from the two lowest $\nu$ bins and would have a significant impact on the quality of the fit, as will be discussed in Sec. \ref{sec:results}.

To minimize initial state effects the observables were published as the double ratio 
\begin{eqnarray}
\label{eq:multiplicity}
R^h_A(\nu,z,Q^{2},p_T^{2})=\frac{\left(\frac{N^h(\nu,z,Q^2,p_T^2)}{N^e(\nu,Q^2)}\right)_A}
{\left(\frac{N^h(\nu,z,Q^2,p_T^2)}{N^e(\nu,Q^2)}\right)_{d}}\, .
\end{eqnarray}
$N^h(\nu,Q^2,z,p_T^2)$ and $N^e(\nu,Q^2)$ are the number of hadrons of type $h$ produced in SIDIS and of inclusive leptons in DIS, respectively. In this double ratio the corrections from nPDFs are negligible to a very good approximation \cite{Sassot:2009sh} and therefore not included in the present work. Only isoscalar effects that take into account the difference between the number of protons and neutrons in the nuclei are included. No significant change is foreseen in the shape of the nFFs nor in the quality of the fit from their inclusion. As the data are given in bins of the kinematic variables one should in principle integrate over the bins. However the mean value for the different bins are shown to be very close to the corresponding averages \cite{Airapetian:2007vu} and those were considered in the computation.

As mentioned above, in the present work only the pion data are studied. The same observable for $\pi^{+}$ has been measured by CLAS \cite{Hakobyan:2008kua} and a brief comparison with the preliminary results is presented in Subsec. \ref{sec:results_non_fitted_data_CLAS}.

\subsection{\label{sec:hadrodata}Hadroproduction}

Single-inclusive identified hadron yields in d+Au collisions at RHIC and p+Pb at the LHC are a useful tool to study cold nuclear matter effects that will contribute to identify genuine signatures of the QGP. Since the previous analysis \cite{Sassot:2009sh} new data from both colliders have become available.

The measured observables are either the cross-sections or the invariant yields. They are related through
\begin{eqnarray}
\label{eq:cross}
\frac{1}{\sigma_{inel}}E\frac{d^{3}\sigma}{dp^{3}}=\frac{1}{N_{ev}2\pi p_{T}}\frac{d^{2}N}{dydp_{T}} \, ,
\end{eqnarray}
where $\sigma_{inel}$ is the total inelastic cross-section which at RHIC is effectively given by the non-single diffractive component. The invariant cross-section as a function of the $p_{T}$ of the outgoing hadron falls various orders of magnitude and it is not easy to see the full extent of the nuclear effects in the differential yields. Instead, the data are usually given in a more appealing form 
\begin{eqnarray}
\label{eq:rdapa}
R_{p(d)A}=\frac{1/N_{ev}d^{2}N_{p(d)A}/dydp_{T}}{\langle N_{coll}\rangle/\sigma^{inel}_{pp}d\sigma_{pp}/dydp_{T}} \, ,
\end{eqnarray}
with $\langle N_{coll}\rangle$ the average number of binary nucleon-nucleon collisions. $\langle T_{p(d)A}\rangle =\langle N_{coll}\rangle/\sigma^{inel}_{pp}$ is the so called nuclear overlap function, obtained through calculations using the Glauber model \cite{Adler:2006wg,Adams:2003qm,Grebenyuk:2007pja}. Further information can be obtained by separating the data into centrality classes, theorized to be related to the geometry of the collision. 

Strictly speaking, in pQCD it is not possible to compute $R_{p(d)A}$ as defined, as neither the nuclear overlap nor the total inelastic cross-sections can be calculated. The latter has to be determined through MC simulations which requires a significant amount of work, so that they are only available for p+p collisions at RHIC and the LHC, and for d+Au at RHIC. The closest thing to quantify the nuclear effects with respect to a proton reference is the ratio of the differential cross-sections in p(d)+A to p+p: 
\begin{equation}
\label{eq:rhicratio}
R^h_{\sigma}(A,p_T)\equiv \frac{1}{2\,A}
\frac{\left. E\,d^3\sigma^h/dp^3\right|_{dA}}
{\left. E\,d^3\sigma^h/dp^3\right|_{pp}}\, .
\end{equation}
Going from $R^{h}_{\sigma}$ to $R_{p(d)A}$ is possible only if $\sigma^{inel}_{pp}$, $\sigma^{inel}_{p(d)A}$ and $\langle T_{p(d)A}\rangle$ are given.

%At contrast with SIDIS, where the PDFs and FFs involved depend on the measured values of the kinematic variables (or have to be integrated over concrete bins at worst), the hadroproduction involves the integration over the full momentum fractions of the three partons (two initial, one outgoing). 
By comparison with SIDIS, at the same order of accuracy the hadroproduction process involves one extra convolution (two initial for the initial partons, one for the outgoing one) and has several more contributing channels. This is made clear by the fact that the pattern of the ratio is not simple, depends on the $p_{T}$ and includes both suppression and enhancement \cite{Adler:2006wg,Adams:2003qm,Grebenyuk:2007pja}. Disentangling how much of the effect comes from the initial state and how much from the final state is not easy as the data can be equally well described by including nFFs \cite{deFlorian:2011fp} or just considering nPDFs \cite{Eskola:2016oht,AbdulKhalek:2020yuc}. The inclusion of the former does provide a slightly better description of the RHIC data, but considering that the observable depends on the proton PDFs, nPDFs, vacuum FFs and nFFs used (and the way these distributions were obtained), it is daring at this point to make a definitive statement. Moreover nPDFs usually have as baseline a set of proton PDFs that already contain some nuclear information and adding nFFs would only lead to a triple counting of effects.

Given the situation and despite their importance for constraining the gluon density, in this work the hadroproduction data are not included in the fit. They are however compared in Subsec.\ref{sec:results_non_fitted_data_RHIC} with the corresponding theoretical predictions, with and without using different modern sets of nPDFs. Data from HI collisions are also left out as they are potentially more affected by the QGP.

\section{\label{sec:results}Results}

\subsection{\label{sec:results_nffs}NLO analysis of pion nFFs}

Minimizing the $\chi^{2}$ defined in Eq. (\ref{eq:chi2_real}) and using the shape proposed in Eq. (\ref{eq:params}), the parameters of the best fit with two possible treatments of the uncertainties were obtained. A priori there is no reason to expect the partons to be modified in the same way by the nuclear medium, and $70$ free parameters should be used. However that much freedom might not be needed. Already for the vacuum case, despite the significantly larger amount of data considered, relations were imposed on the parameters \footnote{E.g. the parameters of $\bar{u}$ and $d$ are taken to be equal and, except for a different normalization, also equal to those of $s+\bar{s}$.}. Therefore in this work the nuclear modification was taken to be the same for all quarks. Repeating the present study for the HERMES kaon data should shed light on this issue, considering the lesser quality of the kaon fit in \cite{Sassot:2009sh} w.r.t. the pion one. 

Assuming no flavour separation of the quark sector, a total of $20$ parameters would be needed. However the data lack the power to constrain all these and further simplifications were made. Similarly to \cite{Sassot:2009sh} a toy model with only $4$ parameters is enough to describe the bulk of the data. Namely, all parameters in Eq. (\ref{eq:params}) are set to zero except $\beta_{q,1}=\beta_{g,1}$, $N_{q,1}$, $N_{g,1}$, and their corresponding $A$ dependence set to be equal: $\beta_{q,2}=\beta_{g,2}=N_{q,2}=N_{g,2}$. A common normalization is no longer possible as now the baseline FFs for gluon and quarks have very different shapes. Such a fit gives $\chi^{2}/d.o.f.=0.93$.

%: one for the $\beta_{i,1}$ with $i=q\text{, }g$, one common power for all the $A$ dependence, and two for the normalizations $N_{i,1}$. A common normalization is no longer possible as now the baseline FFs for gluon and quarks have very different shapes. Such a fit gives $\chi^{2}/d.o.f.=0.93$.

The quality of the fit can be further improved by allowing $\gamma_{q,1}=\gamma_{g,1}\neq 0$ and $\delta_{q,1}=\delta_{g,1}\neq 0$, and using separate powers for quarks and gluons, i.e., $c_{i,2}$ are the same for $N$, $\gamma$ and $\delta$ but $c_{q,2}\neq c_{g,2}$.
As the data lack sensitivity to the $z\lesssim 0.15$ region, $\alpha_{i,j}$ were kept fixed to $0$; releasing these constraints does not result in an improvement of the fit. In total $7$ parameters were needed:
\begin{eqnarray*}
N_{q,1}, &N_{q,2}\\N_{g,1}, &N_{g,2}\\
\beta_{q,1}&=\beta_{g,1}\\
\gamma_{q,1}&=\gamma_{g,1}\\
\delta_{q,1}&=\delta_{g,1} \, .
\label{eq:used_params}
\end{eqnarray*}
The numerical values for the best fit can be found in Tab. \ref{tab:params}, with the fitted parameters in boldface. These and the associated nFFs shown in the following correspond to using Poisson and multiplicative scaling for the statistical and systematic uncertainties, respectively. Removing the scaling does not substantially modify the fit, except for the normalization of the gluon which varies $18\%$. 

%%%%%%%%%%%
\begin{table}[htbp]
\caption{\label{tab:params}%
Values of the parameters describing the nuclear modification of the pion FFs for quarks and gluons. The 7 parameters in boldface are the free parameters determined by the fit. See text for details.}
\begin{ruledtabular}
\begin{tabular}{lrr}
\textrm{Parameter}&
\textrm{i=gluon}&
\textrm{i=quark}\\
\hline\\
$N_{i,1}$     & {\bf -0.0262} & {\bf 0.0322} \\
$N_{i,2}$     & {\bf 0.6654}   & {\bf 0.4567  }\\
$\alpha_{i,1}$& 0               & 0 \\
$\alpha_{i,2}$& 0               & 0 \\
$\beta_{i,1}$ & {\bf -0.0148} & -0.0148 \\
$\beta_{i,2}$ & 0.6654         & 0.4567 \\
$\gamma_{i,1}$& {\bf -0.1555}  & -0.1555  \\
$\gamma_{i,2}$& 0.6654         & 0.4567 \\
$\delta_{i,1}$& {\bf -0.0451} & -0.0451 \\
$\delta_{i,2}$& 0.6654         & 0.4567  \\
\end{tabular}
\end{ruledtabular}
\end{table}
%%%%%%%%%%%%%%%%%%%%%%%%%%%%

\begin{figure*}
\includegraphics{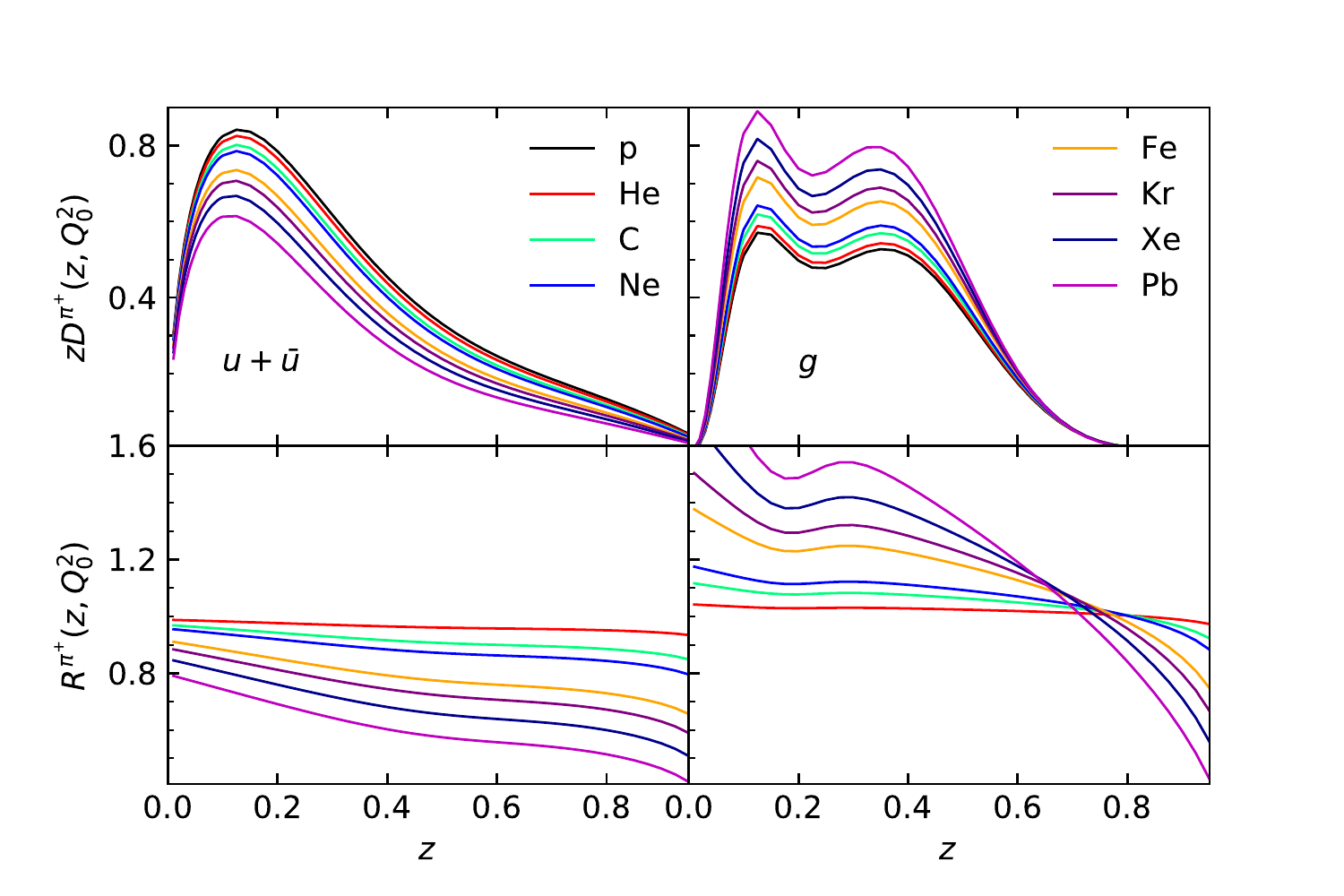}% Here is how to import EPS art
\caption{\label{fig:baseline}Upper row: vacuum DEHSS (solid black) and nuclear LIKEn21 (solid colours) FFs for $u+\bar{u}$ (left) and gluon (right), at initial scale $Q_{0}=1\text{ GeV}$. Lower row: ratio of LIKEn21 to DEHSS.}
\end{figure*}

The comparison of the central values of the nFFs for different nuclei and the vacuum baseline at the initial scale are presented in the upper panels of Fig. \ref{fig:baseline} for $u+\bar{u}$ (left) and the gluon (right). Notice that only some of the nuclei are part of the analysis while the curves for the remaining ones are extrapolations. As in Fig. 7 of \cite{Sassot:2009sh}, in most of the $z$ range the nuclear effects pull the distributions in opposite directions. While the quarks are suppressed with respect to the vacuum (labeled as $p$ in the plot), the gluons are enhanced as shown in the lower right panels of Fig. \ref{fig:baseline}. The trend is inverted for $z\geq 0.8$, a feature also observed in the previous analysis. In contrast, the low $z$ behaviour is very different with almost no suppression for the quarks (lower left) and significant enhancement observed for the gluons. This is quite artificial due to the $\alpha_{i,j}$ parameters that dominate the region being fixed to $0$. Releasing the condition does not affect the quality of the fit but drives the distributions to negative values for $z\lesssim 0.17$. Such a behaviour was observed before (see Fig. 7 of \cite{Sassot:2009sh}) but in the present case it is far more pronounced due to the exclusion of the RHIC data. Given that it can lead to potentially non physical nFFs, the setting $\alpha_{i,j}=0$ was kept. It is however relevant to stress that, due to not including data with $z\le 0.17$, the extrapolation is in principle not reliable in that region. 

\subsection{\label{sec:results_fitted_data}Comparison to fitted data}

\begin{figure*}
\includegraphics{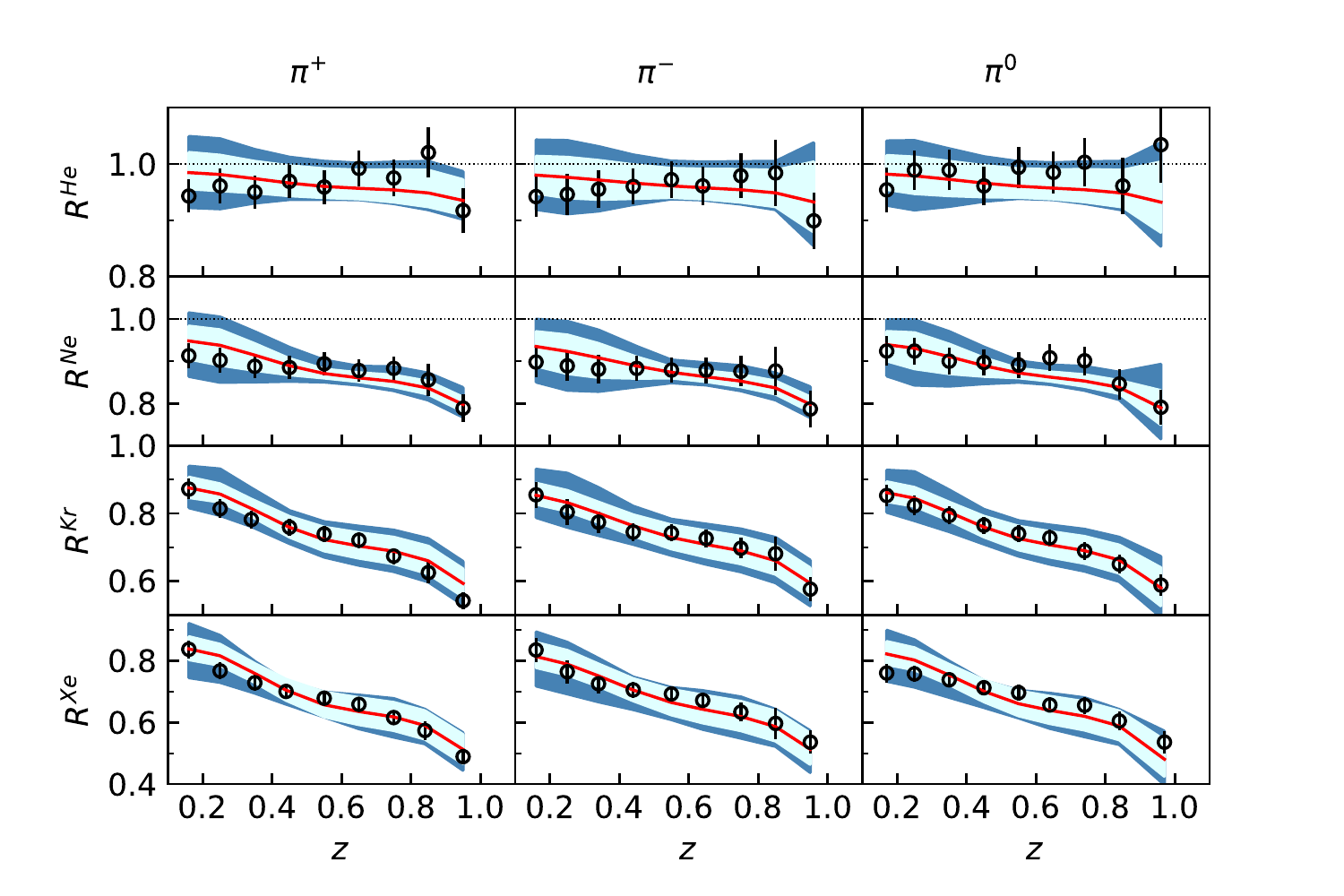}% Here is how to import EPS art
\caption{\label{fig:dep_z}HERMES nuclear SIDIS data for the double ratio and the current fit for $\pi^{+}$ (left column), $\pi^{-}$ (central column) and $\pi^{0}$ (right column) as a function of $z$. Each point corresponds to a different bin in $z$. The inner light (outer dark) band corresponds to the $68\%$CL ($90\%$CL). The statistical and systematic uncertainties are added in quadrature.}
\end{figure*}

\begin{figure*}
\includegraphics{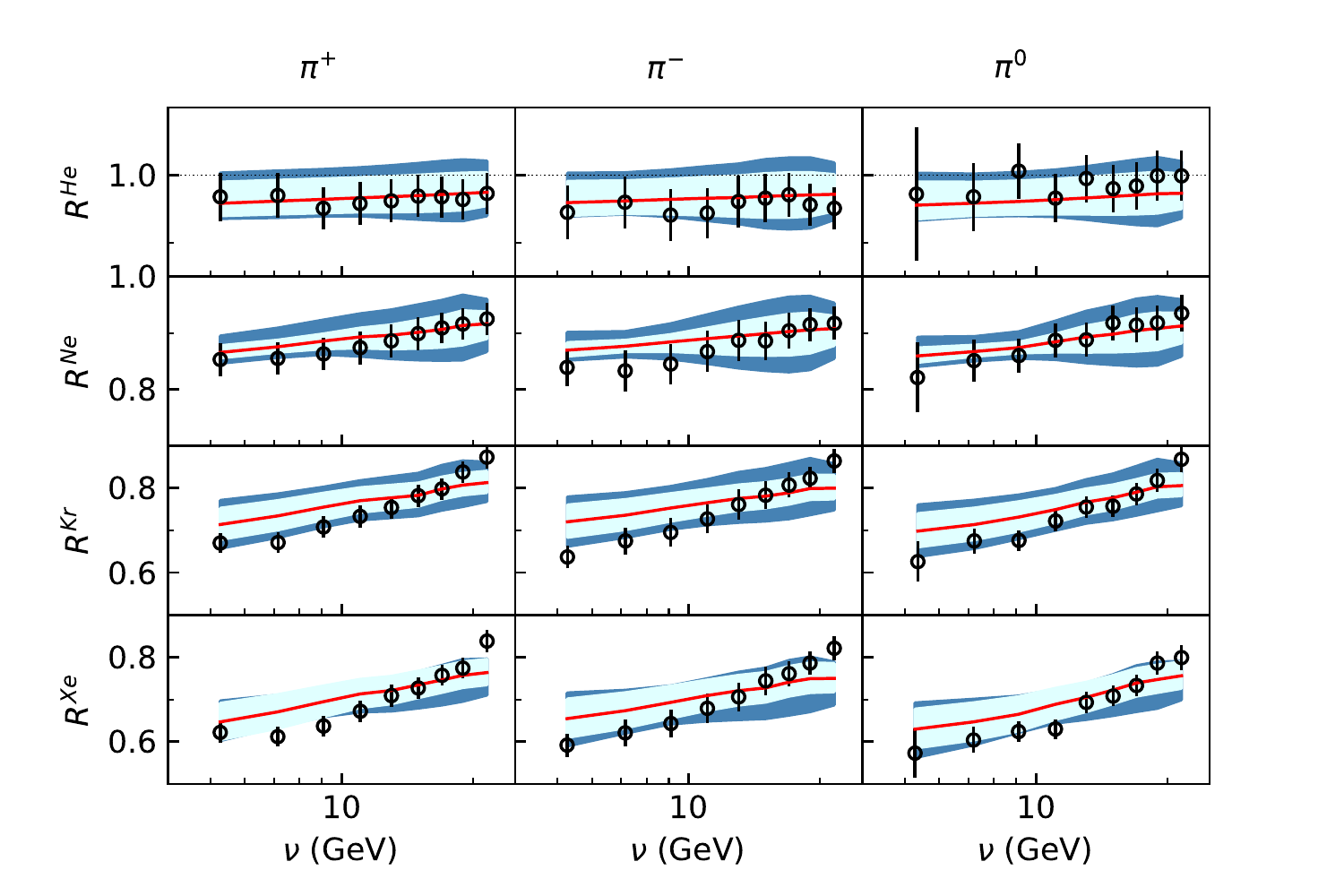}% Here is how to import EPS art
\caption{\label{fig:dep_nu} Same as Fig. \ref{fig:dep_z} but now as a function of $\nu$. The two lowest $\nu$ bins are not within the invariant mass cut $W^{2}$ applied in the proton PDF set used as reference.}
\end{figure*}

The measured multiplicity ratios $R_A^{\pi}$ for pions \cite{Airapetian:2007vu} present a significant dependence in the $z$ and $x$/$\nu$ ranges explored, as can be seen in Figs. \ref{fig:dep_z} and \ref{fig:dep_nu}, respectively. In both cases the (red) curve gives the result of the best fit, the (lighter) inner bands correspond to the $68\%$CL, and the (dark) outer band represents the $90\%$CL. The dependence on $Q^{2}$ is mild and thus not shown here, despite the data being included in and well reproduced by the fit.

As mentioned above, the PDFs taken as reference is MMHT2014, and no nuclear effects were considered beyond the modification for non isoscalarity. While in principle a more reliable result should be obtained using nPDFs, the combination of the current settings in xFitter for the use of initial state nuclear modifications and the available sets of nPDFs in LHAPDF excludes the possibility of using a set determined in the same heavy-flavour scheme as the proton PDFs taken as baseline. Moreover it was shown in \cite{Sassot:2009sh} that no sizable difference would be obtained from doing so. For consistency with the baseline distributions, no nuclear effects were applied either for the deuterium reference, i.e. the deuterium was computed as the average of the proton and neutron cross-sections.

The $z$-dependence of the data is in general very well described for all nuclei in the whole measured range, as can be seen from Fig. \ref{fig:dep_z}. The details of the contribution to the $\chi^{2}$ for each hadron type an medium can be found in Tab. \ref{tab:chi2} for the fits with ($\chi^{2}$) and without ($\chi^{2}_{ns}$) scaling. Comparing with the previous study, there is a reduction of the $\chi^{2}$ of $\sim 8$ units for each charged pion. The $\pi^{0}$ result is dominated by the lowest bin in Xe, and the $\chi^{2}$ of the $z$-dependence experiences a slight increase. As the current work has half the number of free parameters, the effect is most likely a consequence of the greater flexibility of the baseline vacuum FFs.

The $\nu$-dependence, shown in Fig. \ref{fig:dep_nu}, is not reproduced with the same degree of accuracy. While the fit does have a considerable decrease in the $\chi^{2}$ with respect to \cite{Sassot:2009sh} ($\sim 22$, $\sim 14$ and $\sim 9$ units for $\pi^{+}$, $\pi^{-}$ and $\pi^{0}$, respectively), the overall description is not adequate for the heaviest nuclei. In the low $\nu$ region the fit systematically overshoots the data, a trend seen before in this and other approaches, see e.g. Fig. 3 of \cite{Li:2020zbk}. The main contribution to the $\chi^{2}$ here is due to the low $\nu$ (high $x$) region; in particular the quality of the fit could be improved by removing the two lowest $\nu$ bins. In any case these would be discarded if the same kinematic cuts as for the proton PDFs were to be imposed, as their $Q^{2}$ is low ($Q^{2}\le 3\text{ GeV}^{2}$). In this work such a cut was rejected to avoid further reducing the data set, and to highlight the interplay between initial and final distributions. As shown in \cite{Aschenauer:2019kzf,Sato:2019yez} data from SIDIS can significantly alter the shape of the disfavoured parton distributions and thus nuclear SIDIS data have the potential to influence the determination of nPDFs. 

The dependence on $Q^{2}$ in the measured range ($1\text{ GeV}^{2} \le Q^{2} \le 10\text{ GeV}^{2}$) is very mild. The data, roughly one third of the total number of points, contribute to $14\%$ of the $\chi^{2}$.
%
%Table~\ref{tab:chi2},%
\begin{table}[htbp]%The best place to locate the table environment is directly after its first reference in text
\caption{\label{tab:chi2}%
SIDIS data from HERMES \cite{Airapetian:2007vu} included in LIKEn21 nFFs and $\chi^{2}$ of the fit for each nucleus, pion charge and kinematic dependence. For comparison the $\chi^{2}_{ns}$ corresponding to fitting without rescaling the uncorrelated uncertainties is shown. The numerical values are rounded to two decimal positions.}
\begin{ruledtabular}
\begin{tabular}{ccccrr}
\textrm{Nucleus}&
\textrm{Hadron}&
\textrm{Dep.}&
\textrm{\# points}&
\textrm{$\chi^{2}$}&
\textrm{$\chi^{2}_{ns}$}\\
\colrule
$^{4}$He &$\pi^{+}$ & $z$     & 9 & 7.70 & 7.45\\
         &          & $x$     & 9 & 0.52 & 0.53 \\
         &          & $Q^{2}$ & 8 & 1.32 & 1.31\\
         &$\pi^{-}$ & $z$     & 9 & 3.32 & 3.46\\
         &          & $x$     & 9 & 1.48 & 1.55\\
         &          & $Q^{2}$ & 8 & 1.56 & 1.61\\
         &$\pi^{0}$ & $z$     & 9 & 6.25 & 5.75\\
         &          & $x$     & 9 & 3.38 & 3.19\\
         &          & $Q^{2}$ & 8 & 2.72 & 2.56\\
\colrule
Ne &$\pi^{+}$ & $z$     & 9 & 6.13 & 6.25\\
   &          & $x$     & 9 & 1.89 & 1.80\\
   &          & $Q^{2}$ & 8 & 0.98 & 0.92\\
   &$\pi^{-}$ & $z$     & 9 & 3.76 & 3.87\\
   &          & $x$     & 9 & 4.01 & 4.20\\
   &          & $Q^{2}$ & 8 & 0.82 & 0.83\\
   &$\pi^{0}$ & $z$     & 9 & 5.33 & 5.02\\
   &          & $x$     & 9 & 1.93 & 1.91\\
   &          & $Q^{2}$ & 8 & 3.46 & 3.29\\
\colrule
Kr &$\pi^{+}$ & $z$     & 9 & 9.70 & 10.26\\
   &          & $x$     & 9 & 22.04 & 21.80\\
   &          & $Q^{2}$ & 8 & 1.82 & 2.18\\
   &$\pi^{-}$ & $z$     & 9 & 2.85 & 3.01\\
   &          & $x$     & 9 & 21.34 & 22.69\\
   &          & $Q^{2}$ & 8 & 2.23 & 2.43\\
   &$\pi^{0}$ & $z$     & 9 & 2.14 & 2.26\\
   &          & $x$     & 9 & 15.02 & 14.62\\
   &          & $Q^{2}$ & 8 &  4.72 & 5.09\\
\colrule
Xe &$\pi^{+}$ & $z$     & 9 & 7.36 & 7.68\\
   &          & $x$     & 9 & 23.31 & 22.42\\
   &          & $Q^{2}$ & 8 & 2.77 & 3.38\\
   &$\pi^{-}$ & $z$     & 9 & 4.71 & 5.03\\
   &          & $x$     & 9 & 19.74 & 19.43\\
   &          & $Q^{2}$ & 8 & 4.56 & 5.06\\
   &$\pi^{0}$ & $z$     & 9 & 14.34 & 14.85\\
   &          & $x$     & 9 & 15.51 & 15.26\\
   &          & $Q^{2}$ & 8 & 5.19 & 5.90\\
\colrule
Total & & & 312 & 236.73 &239.11 \\
\end{tabular}
\end{ruledtabular}
\end{table}
Overall, the new fit results in a reduction of roughly $110$ units of $\chi^{2}$, giving for the SIDIS multiplicities a total $\chi^{2}/d.o.f=0.776$ ($\chi^{2}_{ns}/d.o.f.=0.784$), quite lower than the previously obtained values. 
\subsection{\label{sec:results_non_fitted_data_CLAS}Comparison to non-fitted data: CLAS}
As mentioned in Sec. \ref{sec:intro}, preliminary data from CLAS using a $5$ GeV beam are also available \cite{Hakobyan:2008kua}. The observable is the double ratio of Eq. (\ref{eq:multiplicity}) for $\pi^{+}$ and C, Fe, and Pb as a function of $\nu$, $Q^{2}$, $z$ and $p_{T}$. Due to the lower beam energy the kinematic space spanned by the data is far more restricted than in the case of HERMES. However these data can be very useful in constraining the final state effects in SIDIS off nuclei. First, the narrow binning diminishes uncertainties generated by not integrating over the whole kinematic bins. Second, the data have very high statistical precision (no quantitative systematic uncertainties are given) which is key for better constraining the parameters of the fit and for flavour decomposition. Third, the nuclei used fill the gaps missing for a better determination of the $A$ dependence, particularly for Pb. The data, including also $\pi^{-}$, are currently under analysis for final publication. In the meantime it is worthwhile to see how adequate (or not) is the prediction of LIKEn21. 

\begin{figure*}
\includegraphics{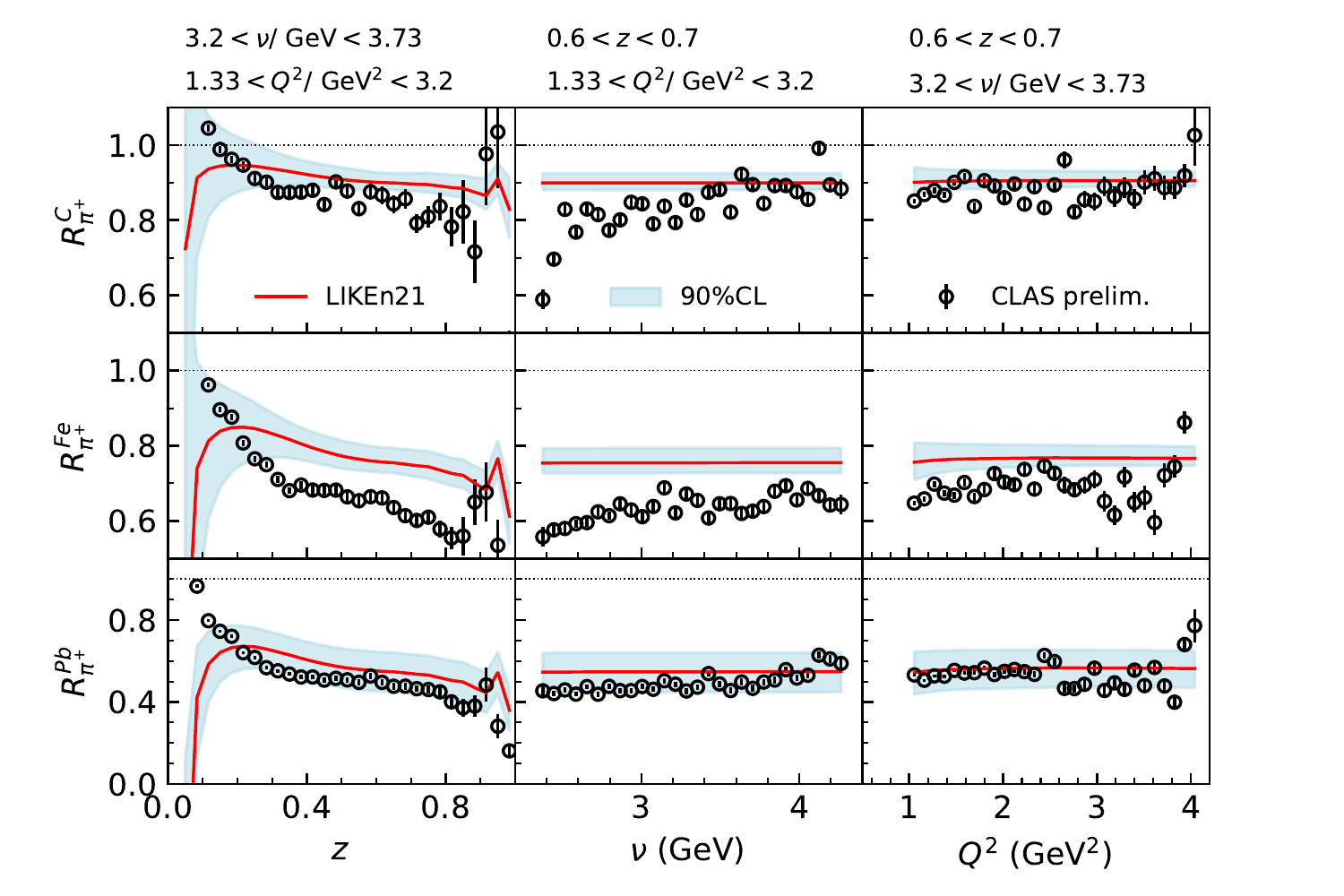}% Here is how to import EPS art
\caption{\label{fig:CLAS_prelim} Preliminary multiplicities measured at CLAS \cite{Hakobyan:2008kua} for C, Fe and Pb as a function of $z$, $\nu$ and $Q^{2}$, and comparison with the prediction using LIKEn21. The band corresponds to the $90\%$CL. For each kinematic dependence the top labels indicate the bins in the other two variables. The bins shown here were selected at random and do not reflect the best nor the worst agreement. The errors bars are only statistical. }
\end{figure*}

In Fig. \ref{fig:CLAS_prelim} a fraction of the data are presented for all three nuclei as a function of $z$, $\nu$ and $Q^{2}$ (left, centre and right columns, respectively). The (red) curve is the central prediction using LIKEn21 and the band corresponds to the $90\%$ CL. 
Setting aside the $p_{T}$ dependent data, there are a total of $700$ points  of which only a randomly selected handful are shown. At first glance it is clear that the $\chi^{2}/N_{pts}$ is not close to unity. However the trend of the data is reproduced, with the largest discrepancies found for Fe. This is unexpected as one would assume that, having Ne and Kr in the fit, the intermediate nuclei would be more or less on point. Surprisingly the Pb data are remarkably well reproduced despite the extrapolation to a much heavier nucleus than Xe. While it is not possible to be conclusive, the CLAS data appears to have a significant constraining power over the nFFs and, when finally published, an update of LIKEn21 will be needed.

\subsection{\label{sec:results_non_fitted_data_RHIC}Comparison to non-fitted data: RHIC}

\begin{figure*}
\includegraphics{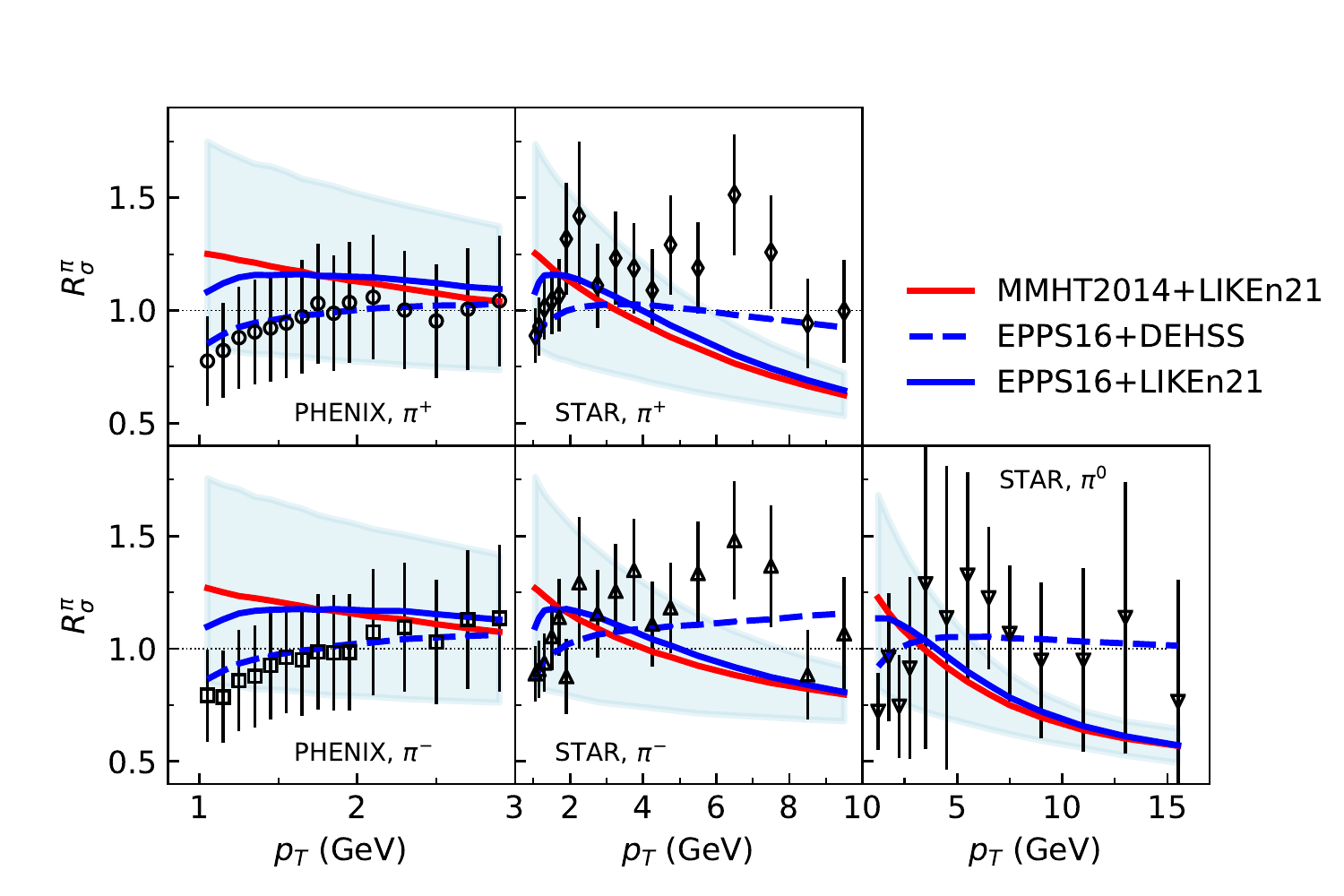}% Here is how to import EPS art
\caption{\label{fig:RHIC} Charged and neutral single pion production at PHENIX (left panels) and STAR. The solid red curve and the band are the central prediction and uncertainty of LIKEn21. The dashed and solid blue curves are the predictions using EPPS16 with DEHSS and LIKEn21, respectively. Using another set of nPDFs, such as nCTEQ15WZ, gives similar results.} 
\end{figure*}

As mentioned above, the pion production in d+Au collisions at RHIC have not been included in the fit due to the difficulty in disentangling initial and final state nuclear effects, and the further computational complication of multiple convolutions. However they played a crucial role for constraining the gluon nFF in \cite{Sassot:2009sh} and, as their removal readily impacts on the shape of that distribution, it is worthwhile to re-examine them under LIKEn21. 
Fig. \ref{fig:RHIC} shows the comparison of the STAR \cite{Abelev:2009hx,Adams:2006nd} and PHENIX \cite{Adare:2013esx} data with the theoretical prediction using LIKEn21 with MMHT2014 proton PDFs (solid red), and the $90\%$CL band for LIKEn21. The dashed and solid blue curves correspond to the predictions using the nPDFs set of EPPS16 \cite{Eskola:2016oht} together with DEHSS and LIKEn21, respectively.
The neutral pion data here (lower right plot) correspond only to the published results from STAR \cite{Abelev:2009hx} which supersede the -previously used- preliminary results of \cite{Grebenyuk:2007pja}. More recent data on charged pions from PHENIX are now also presented in the two leftmost panels of Fig. \ref{fig:RHIC}. In all cases the (overestimated) errors are given as the sum in quadrature of the systematic, statistical and normalization uncertainties.

For $1\text{ GeV}\leq p_T \leq 2\text{ GeV}$ the nFFs present a significant enhancement and coupled with proton PDFs (solid red line) completely miss the suppression in the data. nPDFs provide in that region the needed low $p_{T}$ suppression, as can be seen in the solid blue curve, though the prediction remains well above the data. It is noteworthy to mention that in DEHSS a $5$ GeV cut on $p_{T}$ had to be included to simultaneously fit RHIC and ALICE data. Using it removes all the PHENIX data and almost all from STAR. 

As one moves to the higher $p_{T}$ values, the separation between the solid red and blue curves starts to vanish and the main difference is determined by the nFFs instead. For $p_{T}> 5\text{ GeV}$ LIKEn21 gives a too sharp suppression by comparison with DEHSS (blue dashed curve) and the agreement with the data worsens significantly. 

Considering, however, that RHIC data are fitted in EPPS16 but not in LIKEn21, and their large uncertainties, it is not possible to state beyond reasonable doubt that nPDFs are the only modification needed to describe single particle production at RHIC. Moreover, the RHIC data have already been successfully included in \cite{Sassot:2009sh} and in the nPDFs extraction of \cite{deFlorian:2011fp}, with a $\sim 20\%$ reduction of the $\chi^{2}$ w.r.t. not including final state effects. Furthermore, results from HI collisions suggest that deviations from the proton/vacuum case due to the presence of the nuclear medium are not to be neglected. One could then conclude that the discussion above oversimplifies the situation, and the involved shape of the nuclear modifications put in the spotlight the entanglement between initial and final state effects.

\section{\label{sec:summary}Summary}

A new analysis of medium modified FFs (LIKEn21) based on the vacuum FFs of DEHSS and using the SIDIS pion data measured at HERMES was presented. This work does not aim at explaining the origin of the in medium modifications, but rather to provide a tool for future predictions. 

Following a pioneering work, the nuclear modifications were fitted with a simple ansatz and only 7 parameters were required. The significant improvement in the quality of the fit reflects the increased flexibility of the baseline FFs. Despite using a more reduced data set and half the amount of parameters, in LIKEn21 many characteristics of the original work are reproduced, in particular for the quark sector. The gluon remains severely unconstrained due to the difficulty in efficiently including single hadron production data into the fit. Comparison with preliminary data from CLAS shows signs of constraining power for all partons and possibly the need for further flavour separation, though the final results should be first analysed.

The work was done expanding the xFitter PDF tool to compute also SIA and SIDIS, and incorporating independent parametrizations for the (n)FFs. Whether or not the effects encoded in LIKEn21 will be seen at larger energies at the EIC is yet to be found out, but the changes incorporated will in any case be useful for simultaneous fits of initial and final state distributions.

\begin{acknowledgments}
The author is grateful to M. Walt for her help with the nPDFs modifications in xFitter, and to R. Sassot and M. Epele for enlightening discussions about the DEHSS FFs. C. Andr\'es and A. Sch{\"a}fer are warmly thanked for reading the manuscript and providing constructive comments. Special thanks to I. Zurita, and E., K. and L. Carreira for naming LIKEn21. This work is dedicated to the memory of I. Zurita who left too early and will be sorely missed. The author acknowledges support from the Deutsche Forschungs- gemeinschaft (DFG, German Research Foundation) - Research Unit FOR 2926, grant number 409651613.  
\end{acknowledgments}

\bibliography{main}% Produces the bibliography via BibTeX.

\end{document}